\documentclass{aa}  

\usepackage{graphicx}
\usepackage{color,soul}
\usepackage{txfonts}

\usepackage{natbib}
\newcommand{\sunrise}{\textsc{Sunrise}}
\newcommand{\carcsec}{$\mbox{.\hspace{-0.5ex}}^{\prime\prime}$}
 
\begin{document}

   \title{\textbf{Power spectrum of turbulent convection in the solar photosphere }}

\author{
L.~Yelles~Chaouche\inst{\ref{inst1}} \and
R.~H.~Cameron\inst{\ref{inst2}} \and
S.~K.~Solanki\inst{\ref{inst2}, \ref{inst3}} \and
T.~L.~Riethm\"uller\inst{\ref{inst2}} \and
L. S.~Anusha\inst{\ref{inst2}} \and
V.~Witzke\inst{\ref{inst2}} \and
A.I.~Shapiro\inst{\ref{inst2}} \and
P.~Barthol\inst{\ref{inst2}} \and
A.~Gandorfer\inst{\ref{inst2}} \and
L.~Gizon\inst{\ref{inst2}, \ref{inst4}} \and
J.~Hirzberger\inst{\ref{inst2}} \and
M.~van~Noort\inst{\ref{inst2}} \and
J.~Blanco~Rodr\'{\i}guez\inst{\ref{inst5}} \and
J.~C.~Del~Toro~Iniesta\inst{\ref{inst6}} \and
D.~Orozco~Su\'arez\inst{\ref{inst6}} \and
W.~Schmidt\inst{\ref{inst7}} \and
V.~Mart\'{\i}nez Pillet\inst{\ref{inst8}} \and
M.~Kn\"olker\inst{\ref{inst9}} 
}

\institute{
Centre de Recherche en Astronomie, Astrophysique et Geophysique,
     Route de l'Observatoire, BP 63, Algiers, Algeria\label{inst1} 
                \and
Max-Planck-Institut f\"ur Sonnensystemforschung, Justus-von-Liebig-Weg 3, 37077 G\"ottingen, Germany\label{inst2}
\and
School of Space Research, Kyung Hee University, Yongin, Gyeonggi, 446-701, Republic of Korea\label{inst3}
\and
Institut f\"ur Astrophysik, Georg-August-Universit\"at G\"ottingen, Friedrich-Hund-Platz 1, 37077 G\"ottingen, Germany\label{inst4}
\and
Grupo de Astronom\'{\i}a y Ciencias del Espacio, Universidad de Valencia, 46980 Paterna, Valencia, Spain\label{inst5}
\and
Instituto de Astrof\'{\i}sica de Andaluc\'{\i}a (CSIC), Apartado de Correos 3004, 18080 Granada, Spain\label{inst6}
\and
Kiepenheuer-Institut f\"ur Sonnenphysik, Sch\"oneckstr. 6, 79104 Freiburg, Germany\label{inst7}
\and
National Solar Observatory, 3665 Discovery Drive, Boulder, CO 80303, USA\label{inst8}
\and
High Altitude Observatory, National Center for Atmospheric Research,\footnote{The National Center for Atmospheric Research is sponsored by the National Science Foundation.} P.O. Box 3000, Boulder, CO 80307-3000, USA\label{inst9}
}

   \date{today}

  \abstract
   {The solar photosphere provides us with a laboratory for understanding turbulence in a layer where the fundamental processes of transport vary rapidly and a strongly superadiabatic region lies very closely to a subadiabatic layer. 
Our tools for probing the turbulence are high-resolution
  spectropolarimetric observations such as have recently been obtained with the two balloon-borne \sunrise{} missions, and numerical
  simulations. Our aim is to study photospheric turbulence with the help of Fourier power spectra that we compute from observations and simulations. We also attempt to explain some properties of the photospheric overshooting flow with the help of its governing equations and simulations.
We find that quiet-Sun observations and smeared simulations are consistent with each other and exhibit a power-law behavior in the subgranular range of their Doppler velocity power spectra with a power-law index of$~\approx -2$. The unsmeared simulations exhibit a power law that extends over the full range between the integral and Taylor scales with a power-law index of$~\approx -2.25$. The smearing, reminiscent of observational conditions, considerably reduces the extent of the power-law-like portion of the power spectra. This suggests that the limited spatial resolution in some observations might eventually result in larger uncertainties in the estimation of the power-law indices.

The simulated vertical velocity power spectra as a function of height show a rapid change in
  the power-law index (at the subgranular range) from roughly the optical depth unity layer, that is, the solar surface, to $300$~km above it. We propose that 
  the cause of the steepening of the power-law index is the transition from a super- to a subadiabatic region, in which the dominant source of motions is overshooting
  convection.  
A scale-dependent transport of the vertical momentum occurs. At smaller scales, the vertical momentum is more efficiently transported sideways than at larger scales. This results in less vertical velocity power transported upward at small scales than at larger scales and produces a progressively steeper vertical velocity power law below $180$ km. Above this height, the gravity work progressively gains importance at all relevant scales, making the atmosphere progressively more hydrostatic and resulting in a gradually less steep power law.       
Radiative heating and cooling of the plasma is shown to play
  a dominant role in the plasma energetics in this region, which is important in terms of nonadiabatic damping of the convective motions.}


   \maketitle
%

\section{Introduction}

Turbulent motions occur in many astrophysical contexts, such as stellar convective envelopes, the solar wind, the interstellar medium,
and galaxy clusters \citep{brandenburg2011} and can involve a number of physical processes and properties such as radiation, compressibility, stratification, 
and magnetic fields \citep[][]{Brandenburg2013, mininni2011}.
The Sun in particular provides us with the opportunity of observing turbulent motions at scales from below $10^2$~km
(the resolution limit of existing telescopes) to $1.4 \times 10^6$~km (the solar diameter). 

Observations provide us with maps of the line-of-sight velocity and with intensity images, from which the other components of the
velocity can be inferred \citep[e.g.,][]{Hathaway_etal_2000, Hathaway_etal_2015, rieutord2001, rieutord2010,
Goode_etal_2010, roudier2012, Katsukawa2012, Kitiashvili_etal_2013, yelles2014, asensio_etal_2017}. Previous studies have found widely differing power-law
indices for the subgranular range from $-1$ to $-17/3$ \citep[see, e.g.,][]{rieutord2010, Goode_etal_2010}. This large scatter indicates that some observations may not fully resolve the relevant spatial scales, so that they may be missing some of the power
in the short-wavelength part of the subgranular range. 
This highlights the importance of further investigations with a view to better determining the observed power-law index and its implications in terms of plasma properties. 
Previous telescopes stationed above the Earth's atmosphere (e.g., Hinode) have produced consistent seeing-free data that allowed us to study velocity power spectra \citep[see, e.g.,][]{rieutord2010}.
Here we make use of the spectral imaging capabilities of the \sunrise\ balloon-borne observatory, whose core is a $1$m telescope suspended from a
stratospheric balloon that carries it above $99\%\ $of the Earth's atmosphere. This is the largest solar telescope ever to leave the ground and thus
provides a new window for studying solar convective turbulence.

\sunrise\ observations allow us to investigate astrophysical turbulence under the conditions of the highly stratified solar photosphere.
The stratification is superadiabatic
below the solar surface and subadiabatic above it.
The major energy transport mechanism changes from convective beneath the surface to radiative above it.
Filamentary magnetic fields are present,
as are velocities that almost reach the local sound speed. The relative contributions of different physical processes therefore
change rapidly with height. This makes the problem of understanding the turbulent motions both interesting and difficult.
In these circumstances numerical simulations greatly enhance our understanding.  
Such simulations (albeit at much lower Reynolds numbers than for the Sun)
have been used to
address turbulence in the photosphere \citep[e.g.,][]{Stein_Nordlund1998, rieutord2001, Georgobiani2007, brandenburg2011, moll2011, Kitiashvili_etal_2013, yelles2014} at various scales and resolutions for almost $40$ years \citep{Stein1989}.
In this paper we concentrate on the changes in the turbulent spectrum throughout the solar photosphere by comparing observations
and simulations.

\section{Observations and data reduction}

The observations were recorded during the first two flights of the balloon-borne observatory \sunrise{} (\cite{barthol_etal_2011}) in 2009 June
and 2013 June. The balloon flew at an altitude of roughly $36$ km, which minimized image distortions introduced by the Earth's atmosphere.
An overview of the two flights can be found in \cite{solanki2010} and \cite{solanki_2017}.
The observations considered herein were recorded with the Imaging Magnetograph eXperiment \citep[IMaX;][]{martinezpillet2011} and with the Sunrise Filter Imager \citep[SuFI,][]{gandorfer2011}. For more details on the instrumentation, see also \cite{barthol_etal_2011} and \cite{berkefeld2011}. 
IMaX had a field of view of $51\times51$ arcseconds with a
plate scale of $0$\carcsec{}$0545$ per pixel and a spectral resolution of $85$\,m\AA{}, while for SuFI these numbers are $15\times40$ arcseconds, $0$\carcsec{}$0207$ per pixel and $5$ nm. 
IMaX data were recorded with a cadence of $33.3$ s during the first flight and $36.5$ s during the second flight. The exposure time for each of the fours Stokes parameters and at each wavelength position was $250$ ms. 
The cadence in the case of the SuFI $300$ nm series is $7.2$ s and the exposure time was $500$ ms.
We used these data mainly because they have a very high spatial resolution and are of a constant quality that is unaffected by atmospheric turbulence (seeing). This allows us to probe spatial frequencies that have not been commonly accessible before.

We used four time series. The first two were recorded from 00:36 to 00:59~UT and 01:31 to 02:00~UT on 2009 June 9, while the \sunrise{}
telescope pointed to a quiet-Sun region at disk center. 
The third time series was recorded from 23:39 to 23:55~UT on 2013 June 12 when the telescope pointed to the trailing part of
active region AR11768 (heliocentric angle $\mu=0.93$). Line-of-sight velocity maps from the three first data sets were determined by inverting
the Stokes parameters of the Fe I $5250.2$ \AA\ line with the Stokes-profiles-inversion-O-routines code \citep[SPINOR;][]{frutiger2000} as described in \cite{solanki_2017}. The final time series is composed of SuFI
$300$ nm broadband images recorded from 23:39 UT on 2013 June 12 to 00:38 UT on 2013 June 13 at a heliocentric angle $\mu=0.93$.
We used the upper right region of the SuFI field of view, which has some quiet-Sun-like characteristics, but still contains magnetic field concentrations \citep{jafarzadeh_etal_2017}.

\section{Simulations and spectral synthesis}\label{sec:simulations}

We used comprehensive (including compressible magnetohydrodynamics,
radiative transfer in a non-gray approximation, and the effects of partial ionization on the equation of state)
simulations of the solar photosphere and the topmost layers of the solar interior. The simulations were performed with the Max Planck Institute for Solar System
Research/University of Chicago radiation magnetohydrodynamics code \citep[MURaM;][]{Voegler_etal_2005}.
\cite{Riethmueller_etal_2014} showed that a run with an initial $30$G vertical magnetic field was best suited to reproduce the magnetic and brightness characteristics of the quiet Sun. We used a similar run as \cite{Riethmueller_etal_2014}. After the magnetic field was included, the simulation was kept running for several solar hours to reach a statistical stationary state. The simulation box had horizontal dimensions of $(6 \times 6)$ Mm$^{2}$ with a grid size of $(10.42 \times 10.42)$~km$^{2}$.
The box was $1.4$ Mm deep and extended from $-900$ km below the solar surface to $500$ km above it, with a $14$ km vertical grid spacing. The time series used here covers about one hour of solar time. We will refer to this run as the $30$ G MHD run.

In order to compare simulations with IMaX observations, we first computed Stokes profiles from the simulations using the forward-calculation mode of the SPINOR inversion code \citep{solanki1987, frutiger2000}. The computation was made in local thermodynamic equilibrium (LTE) for the IMaX line Fe I 5250.2 \AA\ in each atmospheric column of the simulation cube.  
The first quantity we extracted from the simulations was the Doppler (line-of-sight) velocity corresponding to the Fe I 5250.2 \AA\ line. The Stokes data were degraded before this to reach almost observational conditions (as described in detail in Sect.~\ref{sec:smearing}). 
 
With a view to comparing the SuFI data with simulations, we computed the brightness from a $\approx 10$ nm band in the vicinity of the $300$ nm wavelength. This band contains 1202 identified lines and it therefore takes extremely long when computed line by line. We therefore proceeded by computing the intensity using the opacity distribution function (ODF) approach, in which the opacities of numerous atomic and molecular lines are averaged into specific wavelength bins for a set of temperature, pressure, and microturbulent velocity values \citep[see, e.g.,][pp.625-627]{hubeny_mihalas_2015}. 
The ODF approach is much faster, but accurate enough for our purposes \citep[see the detailed analysis in ][]{Miha2019}. We computed the ODF for solar metallicity with a microturbulence of 2 km s$^{-1}$ using a modified version of the spectrum synthesis program \citep[DFSYNTHE;][]{Castelli_2005_DFSYNTHE, Kurucz_2005_Atlas12_9}. This ODF table was used in the radiative transfer code ATLAS9 \citep{Kurucz_manual_1970, Castelli_ATLAS12_2005} to compute the emergent intensity for the vertical rays of the MURaM box. The ATLAS9 code assumes LTE.

\section{Adjusting simulations to observational conditions}\label{sec:smearing}

In order to adjust the simulations to observational conditions, it is necessary to apply an adequate smearing to the simulations that takes the degradation due to instrumental and observational conditions of IMaX and SuFI into account. We followed a method similar to the methods used in \citet{Riethmueller_etal_2017}, \citet{Riethmueller_etal_2014}, and Ghosh et al. (in prep.).

\subsection{Spectral smearing and resampling}

The spectral point spread function (PSF) of IMaX  as well as the transmission profiles of the SuFI filters were determined experimentally in the laboratory before the launch of \sunrise\  \citep[][]{martinezpillet2011, gandorfer2011,Riethmueller_etal_2014}.
After computing the Stokes profiles from MHD simulations using the SPINOR code
\citep{solanki1987, frutiger2000}, we first convolved them with the IMaX spectral PSF in order to reproduce the spectral resolution of IMaX.
We then resampled the resulting profiles at spectral locations similar to those used in the IMaX instrument: $\pm 80$ m \AA\ and 
$\pm 40$ m \AA\ from line center, and in the continuum at $+227$ m \AA\ . The data were assembled as 2D images each corresponding to a spectral position for a given Stokes parameter (we are mainly interested in Stokes-I here). This format is more adequate for the coming spatial smearing.
The synthetic SuFI data were computed after weighting the contribution of each spectral position by its corresponding part of the transmission profile of the SuFI filter at $300$ nm. Therefore the spectra located at the very center of the $300$ nm filter contribute more to the resulting intensity than the spectra located at the edge of the filter. This inclusion of the SuFI transmission profile was made in the early phases of the ODF spectral synthesis (described in Sect.~\ref{sec:simulations}). Again, the resulting intensity 2D images were stored in view of the next spatial smearing.     
\subsection{Spatial smearing}

The observational data were recorded in phase-diversity (PD) mode and were carefully reconstructed \citep[e.g.,][]{solanki_2017, martinezpillet2011}. The PD measurements allow us to determine the wavefront aberrations of the system. This enables us to determine the spatial PSF and subsequently use it to deconvolve the data accordingly.  The retrieved PD PSF deconvolved data would theoretically have a spatial resolution at almost the diffraction limit of the telescope. Even though some SuFI data reached an extremely hight resolution of $\approx$ 0\carcsec{}1 \citep{gandorfer2011}, the bulk of the data has a somewhat lower resolution. This indicates that some further instrumental effect smears the data. The main candidates for this additional blurring are issues related to the residual pointing (jitter) of the telescope \citep[see other possible reasons in][]{Riethmueller_etal_2017}.
Consequently, in order to compare synthetic Stokes and intensity data with observations, we convolved the synthetic data with a spatial PSF that accounted for the blurring due to the telescope jitter. We chose a PSF made by a 2D Gaussian filter (another choice might be a combination of two Butterworth lowpass filters). By comparing azimuthally averaged Fourier spectra from observations and smeared simulations and comparing the rms quiet-Sun contrast of the two data sets, we found that a PSF with a full width at half maximum (FWHM) = 0\carcsec{}145   matches simulations and IMaX observations best. The best match for the SuFI filter at 300 nm is reached for a FWHM = 0\carcsec{}127.

\subsection{Stray light and noise}
In principle, the stray-light correction can be done by recording 
the solar limb profiles during observations and comparing them with 
those from literature \citep[see, e.g.,][]{dunn1968}. This would allow the stray-light modulation transfer functions (MTF) to be calculated.
Then the Stokes data are corrected for stray light contamination by multiplying them 
with the stray-light MTF in Fourier space. This was the case for the 
first \sunrise\ flight. Because of technical problems in the second flight, it was not possible to apply
this procedure. A possible solution was contaminating
the simulated Stokes data with different levels of stray light (together with the smearing
procedure described above) and compare the properties of the retrieved data with 
observations. This allowed us to determine the optimal stray light that is needed 
to retrieve the intensity contrast of observations and other properties that assess the
validity or quality of the smearing procedure we used  
\citep[see][for a more detailed description]{Riethmueller_etal_2017}.
Finally, noise occurs naturally in observations. We chose to add white noise to the Stokes-I and $300$nm band data
with a level between $1.5 \times 10^-3$ Iq and 
$7 \times 10^{-3}$ Iq  (Iq is the quiet-Sun intensity).

\subsection{p-mode filtering}

We are essentially interested in the subgranular range of scales. Although this range is only marginally affected by p-modes, we proceeded to filter the excess energy related to these acoustic modes.
The p-modes were filtered from the data with
a subsonic filter. We removed modes that were located in the region where
$\omega/k > V_{filter}$ in the ($k,\omega$) diagram. $V_{filter}$ is a chosen cutoff velocity above which modes were filtered. We used $V_{filter} =5$ km/s \citep[see, e.g.,][]{yelles2014}. A simulation box (acting as a cavity) might not contain the same amount of p-modes as observations. If the large-scale part of the power spectra is to be studied in detail, the filtering procedure therefore has to be combined with a careful monitoring of the distribution in the ($k,\omega$) diagram (Ghosh et al. in prep.).

\subsection{Doppler velocity and intensity maps}

\begin{figure}
\includegraphics[width=0.48\textwidth]{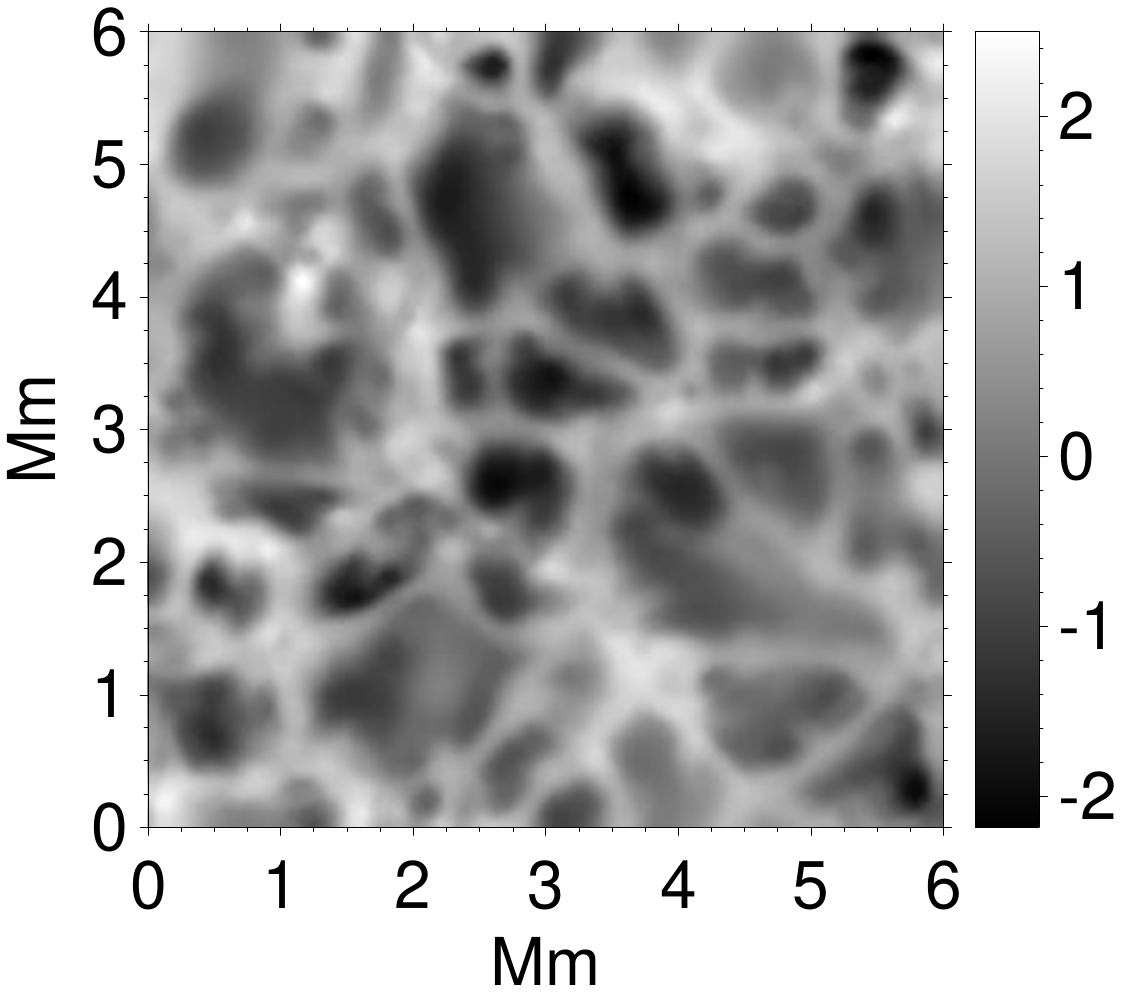}
\includegraphics[width=0.48\textwidth]{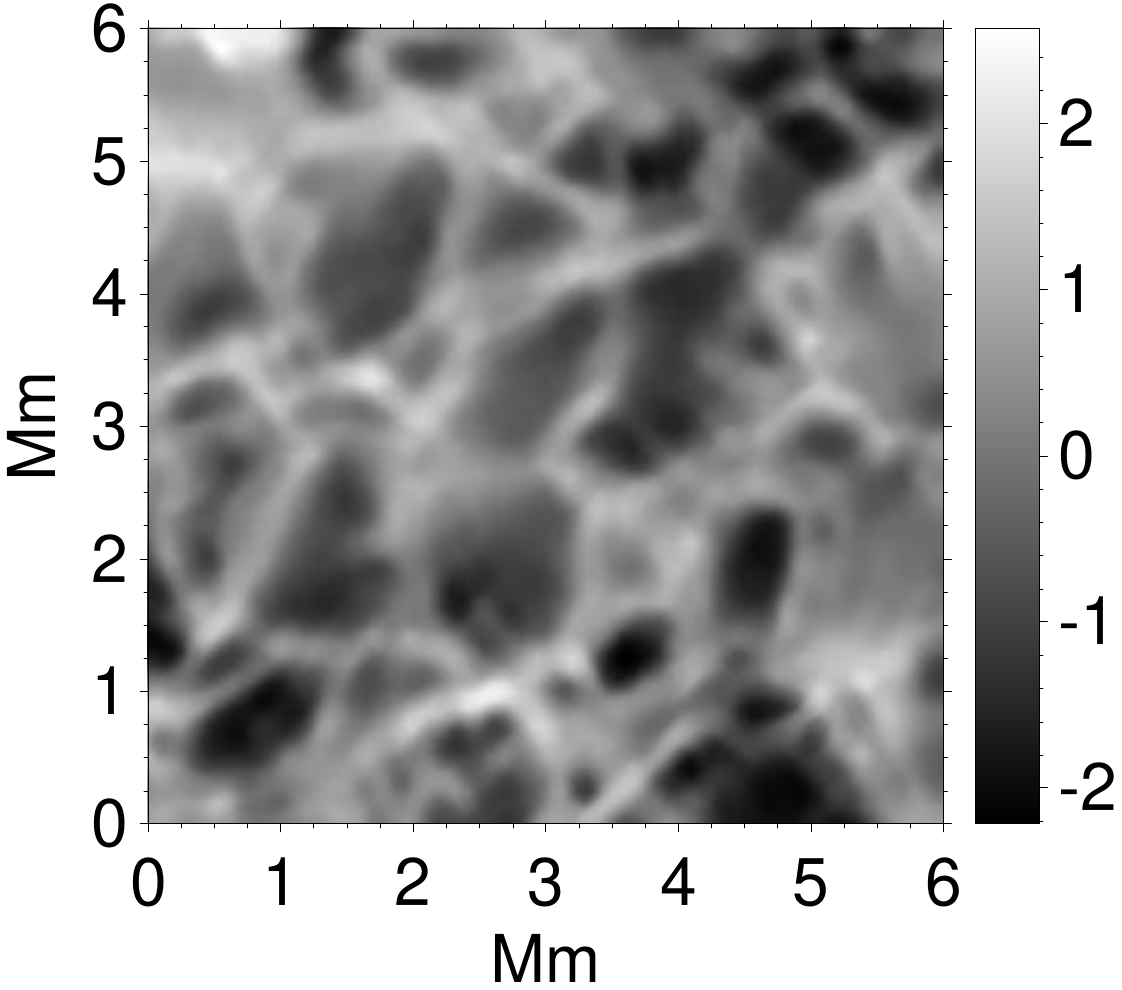}
\caption{Top: Doppler velocity observed by the SUNRISE/IMaX instrument
using the spectral line Fe I 5250.2 \AA. The size of the displayed domain is chosen to be $6 \times 6 Mm^{2}$ for comparison with the simulation domain. Bottom: Doppler velocity from simulations using the same spectral line as for the observations. The displayed Doppler map has been degraded to approach observational conditions. The velocity is expressed in km $s^{-1}$. Positive velocities correspond to downward-directed flows.  
}\label{fig_Doppler}
\end{figure}

After degrading the simulated data to almost observational conditions, we computed Doppler maps to compare them with those obtained by the IMaX instrument. 
The Doppler velocity for observations was obtained with an inversion procedure applied to the Stokes data \citep[see][for more details]{solanki_2017}. The inversion used three nodes along the optical depth scale for the temperature, while the remaining parameters (including the Doppler velocity) were taken to be independent of height. The resulting Doppler velocities for the quiet-Sun are very similar to those obtained by fitting a Gaussian function to the Stokes-I profiles and determining their shift from a reference value. This works especially well for our simulations that are made for quiet-Sun conditions. The mean relative difference in our data between the Doppler maps obtained with the two methods is typically about $1$\%. The difference between the two methods is probably so small because the Stokes data we used are spatially degraded and the spectral line was sampled at only four spectral locations plus the continuum.
For the simulated data, we therefore computed Doppler velocities using the much faster Gaussian fitting method. A sample of the obtained Doppler maps from simulations is displayed in the bottom panel of Fig.~\ref{fig_Doppler}. 
Downflows (associated with positive values of the side bar) are associated with intergranular lanes, and upflows are reminiscent of the granule body.
Observations (upper panel) were sampled from the first IMaX time series taken on 2009 June 9 and represent a quiet-Sun region. In order to be able to appreciate the similarities and differences between observations and simulations, we chose to display a portion of the observational map of the same spatial size as the simulations. Nonetheless, the full field of view of IMaX data spans about $32 \times 32 Mm^2$. Some similarities between the two Doppler maps are evident. A more quantitative comparison is presented in Sect.~\ref{sec:Doppler power spectra }

\begin{figure}
\includegraphics[width=0.48\textwidth]{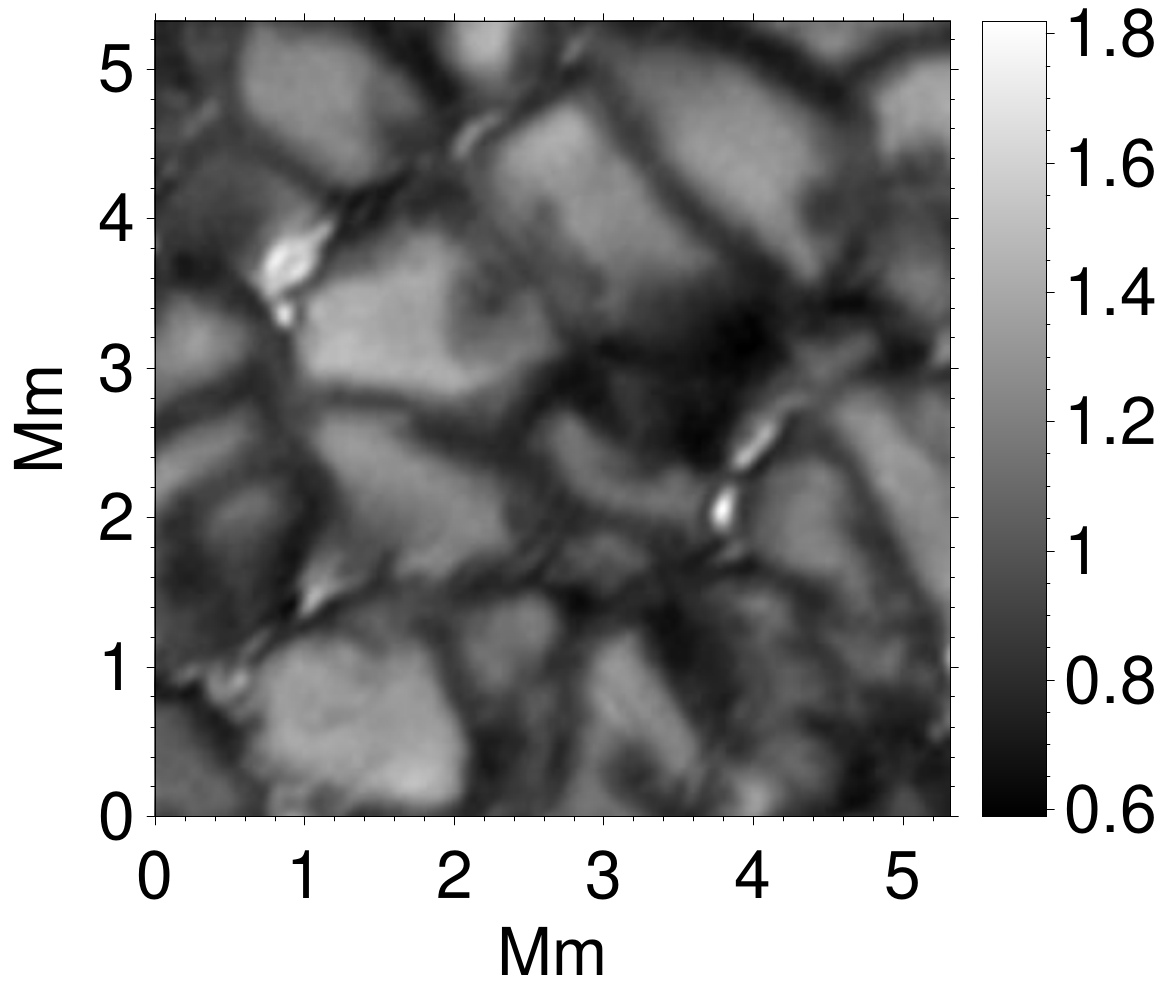}

\includegraphics[width=0.48\textwidth]{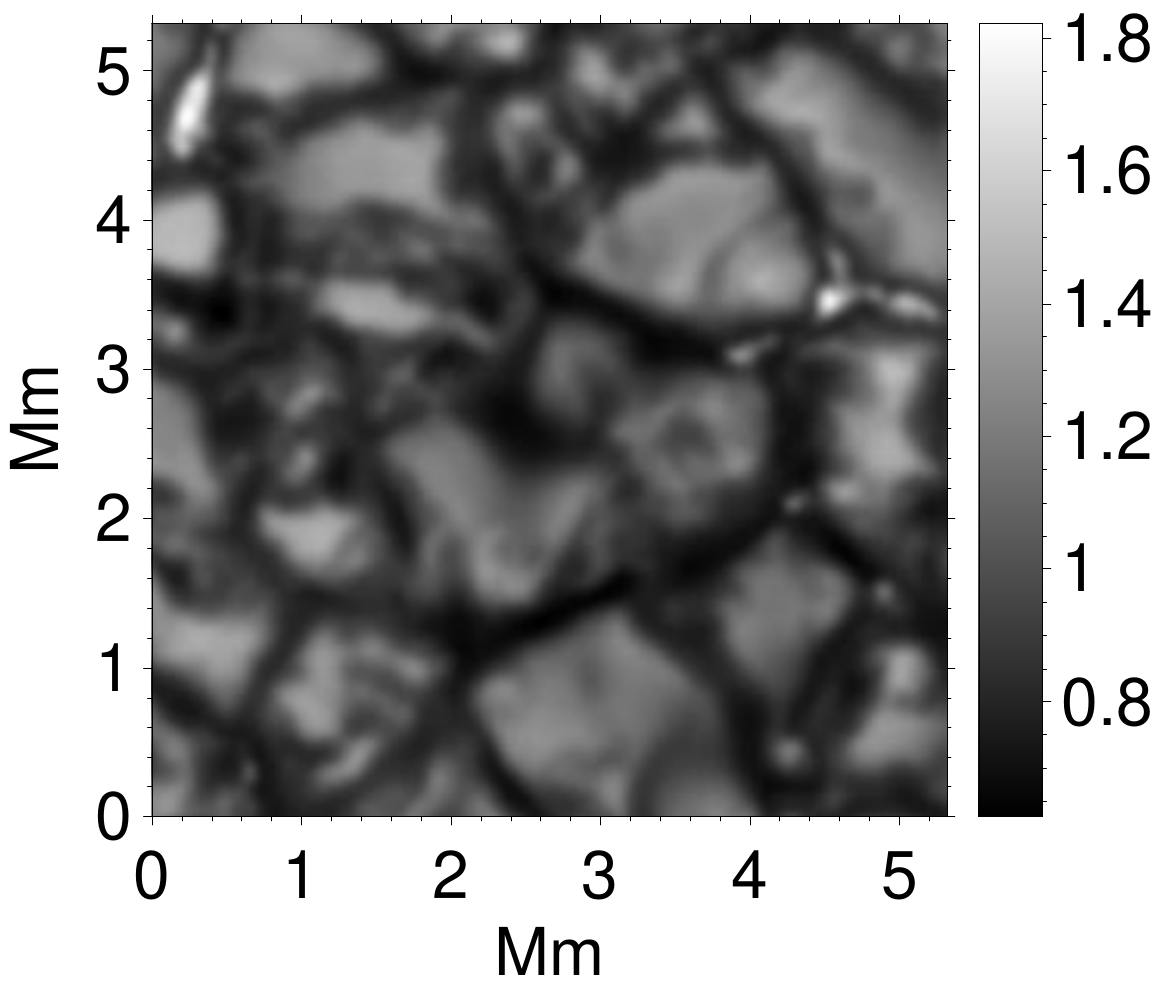}

\caption{Top: Normalized intensity observed by the SUNRISE/SuFI instrument
in the 300 nm band. Bottom: Normalized intensity from simulations in the 300 nm band. The simulated map was degraded to adjust it to observational conditions (as detailed in Sect.~\ref{sec:smearing}).  
}\label{fig_Intensity}
\end{figure}

The second quantity we desired to obtain from the simulations is the brightness (intensity) near the IMaX line and in the $300$ nm band. In the first case, and after smearing the data to almost IMaX conditions, the brightness was extracted from the Stokes-$I$ profile of the Fe I 5250.2 \AA\ in the continuum at an offset of $+227$ m\AA\ from the line center.
In the case of the $300$ nm band, we  applied the smearing procedure relevant for SuFI conditions to the simulated intensity data obtained by ATLAS9 (Sect.~\ref{sec:simulations}). An example of the resulting smeared-intensity maps in the $300$ nm band is shown in the lower panel of Fig.~\ref{fig_Intensity}. The typical spatial resolution of this map is $0.127 \arcsec$. A sample of the observed intensity maps in the $300$ nm band is displayed in the upper panel of Fig.~\ref{fig_Intensity}.

\section{Power spectra: observations and simulations}

We considered power spectra of the Doppler velocity and intensity images. The region we are mostly interested in is located at higher wavenumber (k) than the integral scale (see the definition below). In this region, the power spectra have an approximately power-law-like shape.
Although a power-law regime (at higher k than the integral scale) is commonly associated with
an inertial regime, it is too premature to assume this because the power-law regime in our case spans less than a
decade in k (see, e.g., the discussion in \cite{Rempel_2014} and \cite{Danilovic_etal_2016} for recent references).
In the following, we define a region (in k) in which we study the power-law behavior of the power spectra. This region is defined to lie between the integral and Taylor scales for the un-smeared simulations \citep[see, e.g.,][]{moll2011}.
For the observations and smeared simulations, we considered the resolved part of the power spectra located between the integral scale (or close to it) and a few times the
resolution limit. The integral scale ($L_0$), the location at which energy is injected into the flow,
and the Taylor scale $\lambda_T$, where viscosity effects begin to become significant \citep{batchelor1953, weygand2007}, are given by 

\begin{equation}\label{eq:L_0}
  L_0 = \int_0^{\infty} k^{-1}\, E (k)\,dk / \int_0^{\infty} E(k)\,dk \;,
\end{equation}

and

\begin{equation}\label{eq:lambda}
  \lambda^{2}_{T} = \int_0^{\infty}  E (k)\,dk / \int_0^{\infty} k^{2}\, E(k)\,dk \;,\end{equation}
where $E (k)$ is the power spectrum under consideration, and $k$ is the wavenumber. 

For a given physical quantity (velocity or intensity), which we call $f$ in a given map, the power
spectrum $E (k)$ is defined according to Parseval's theorem as
\begin{equation}\label{eq:powerspectrum}
E (k) =\frac{1}{2}  k\int_0^{2\pi} 
\frac{\left| \strut [{\cal F}(f)](\vec{k})\right|^2}{A}
\;d\theta\;,
\end{equation}
with ${\cal F}(f)$ the 2D Fourier transform of $f$,
$A$ the area of the spatial
domain, and $k$ and $\theta$ the module and orientation angle, respectively,
of the wavevector $\vec{k}$. 
Power spectra are computed for 2D velocity and intensity maps. The computation in the Fourier domain was made over rings of $k\, dk\, d\theta$. This produced an azimuthal averaging to obtain power spectra that only depend on k.

\subsection{Doppler velocity power spectra}\label{sec:Doppler power spectra }

\begin{figure*}
\includegraphics[width=0.95\textwidth]{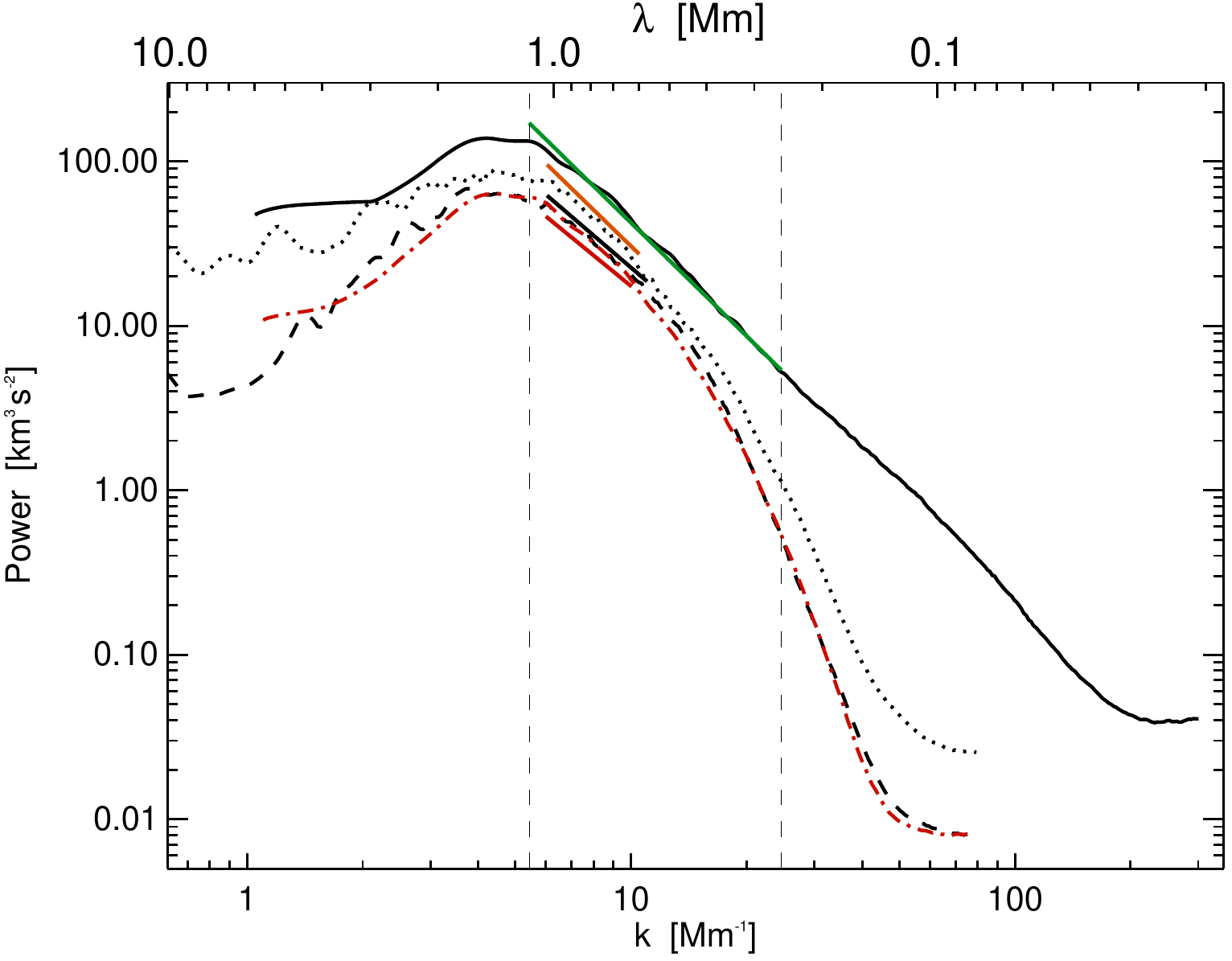}
\caption{Spatial power spectra of the Doppler velocity.
We use the spatial wavenumber ($k=2 \pi/ \lambda$, lower horizontal
axis) and the corresponding wavelength $\lambda$ (upper horizontal axis) as abscissa. 
The dashed black curve corresponds to the power spectrum of the quiet-Sun IMaX time series, and the active-Sun spectrum (second IMaX flight) is represented by the dotted black line. The power spectrum of the Doppler velocity obtained from simulations (from spectral synthesis of the IMaX spectral line) without degradation is represented as a solid black line. The dash-dotted red curve corresponds to the power spectrum of the Doppler velocity from simulations (as the black curve), but this time, we applied spatial and spectral degradation to approach the IMaX observational conditions (as described in Sect.~\ref{sec:smearing}).
The straight green line represents a $k^{-2.25 \pm 0.072}$ power law determined as a fit to the underling black curve. The $\pm$ part represents the uncertainty of the least-squares fit.
The straight orange line is a $k^{-2.24 \pm 0.27}$ power-law fit of the corresponding part of the dotted black curve.
The straight black line corresponds to a fit of the corresponding portion of the dashed black curve and corresponds to $k^{-2.0 \pm 0.22}$.
Finally, the straight red line is a $k^{-1.96 \pm 0.25}$ fit of the dashed red line.
 The two vertical dashed lines indicate the location of the integral and Taylor scales.}\label{fig0}
\end{figure*}

Figure~\ref{fig0} depicts power spectra of the Doppler velocity. 
The plots include observational curves corresponding to the quiet and the active Sun, and also MHD simulations.  
The power spectrum of the Doppler velocity (corresponding to the IMaX spectral line) obtained from the 30G MHD simulations is plotted as a solid black line. For this curve, we did not apply any spectral or spatial smearing. 
%
Fig.~\ref{fig0} shows that after the smearing is applied, the smeared-simulation curve is quite similar to the observational quiet-Sun curve, especially in the subgranular range of scales. This indicates that the simulations are likely to replicate conditions similar to the observed quiet Sun (as far as power spectra are concerned). It is therefore natural to explore the unsmeared power spectrum and assume that it probably represents the quiet Sun to a certain extent without the instrumental degradation inherent to observations. 

Below the integral scale, the unsmeared simulation exhibits a power-law shape with an index of$~-2.25$. To determine this, we made a least-squares fit with a power-law-shaped function. The fitted portion of the curve is located between the integral and Taylor scales. The resulting fit is achieved with an uncertainty of $0.072$.
A small fraction of the degraded simulation power spectrum and the two IMaX spectra exhibits an approximate power-law behavior between $\lambda \approx 0.6$ Mm and $\lambda \approx 1$ Mm. 
The least-squares fitting procedure was applied to the curves in the portion that correspond to the straight lines indicated in Fig.~\ref{fig0}. 
They exhibit indices of $~-2.24 \pm 0.27$, $~-2.0 \pm 0.22$ and $~-1.96 \pm 0.25$ corresponding to the active and the quiet Sun, and the smeared MHD simulations, respectively. 
Other authors have performed similar power spectra analyses on different data sets. Thus \cite{rieutord2010} obtained $-10/3$ and $-17/3$ for the power-law exponent, and \cite{Katsukawa2012} reported $-3.6$. The
power spectrum in Fig.~\ref{fig0} reaches a slope of approximately $-17/3$ only at small wavelengths where the limited resolution of the telescope affects the observations.

The extent of the power-law-shaped part of the smeared simulations (as well as observation) is far smaller than the unsmeared simulations, even though the effective resolution of the data is $\approx$ 0\carcsec{}145. This suggests that at even lower resolution, the extent of the power-law part might be even more reduced. This might result in larger uncertainties in the estimation of the power-law indices.
Figure~\ref{fig0} suggests that the spatial resolution of the data is an important criterion to consider when we wish to observationally explore photospheric velocity power spectra with a more extended power-law part.

\subsection{Intensity power spectra}

\begin{figure*}
\includegraphics[width=0.95\textwidth]{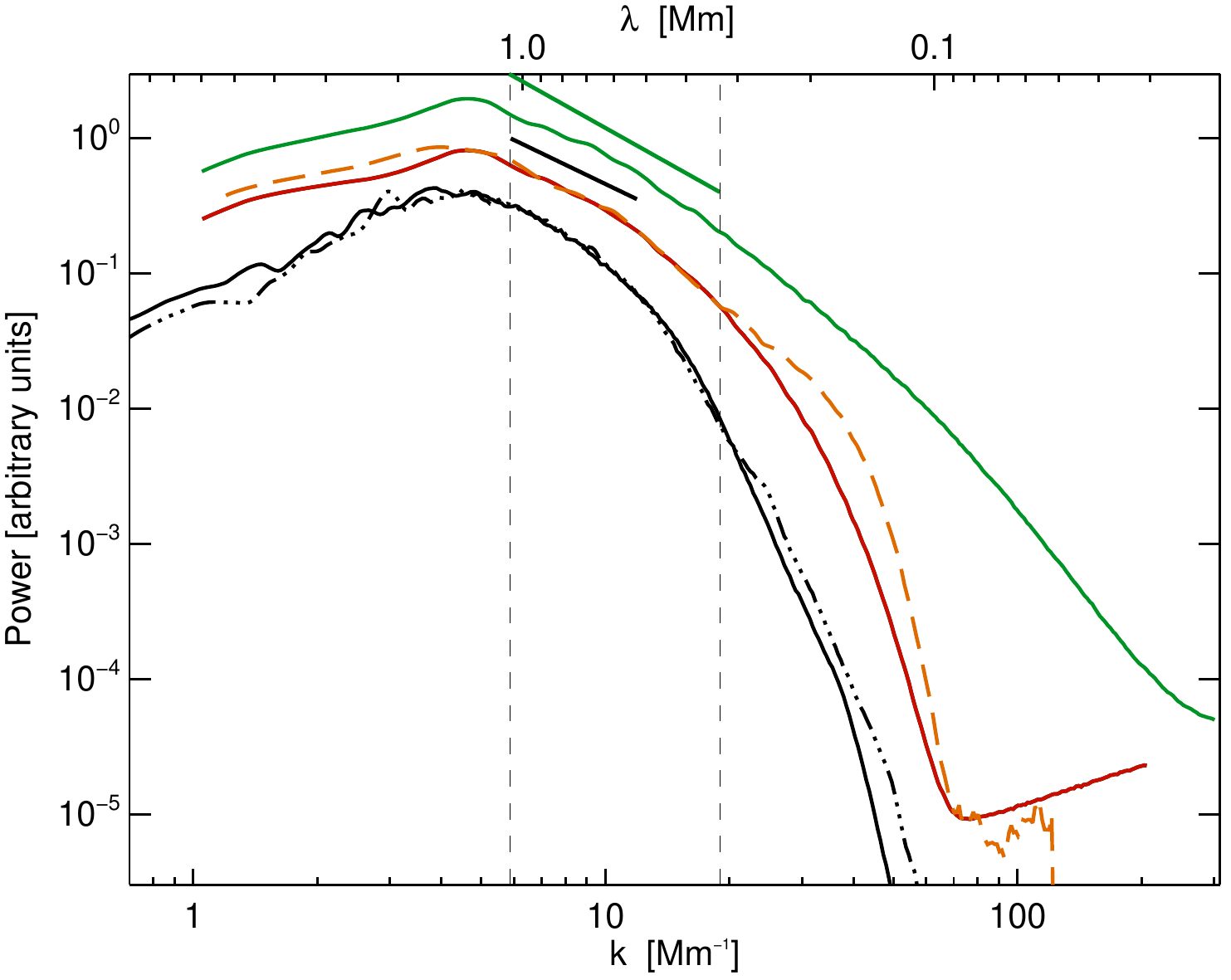}
\caption{Power spectra of the intensity maps. Spectra for the two quiet-Sun IMaX time series are plotted as solid and dash-triple-dotted black lines. The dashed orange curve corresponds to the power spectrum derived from data acquired by the SuFI instrument. The solid red curve is obtained from the simulations through an ODF spectral synthesis of the full SuFI filter at the 300 nm band and applying the smearing procedure presented in Sect.~\ref{sec:smearing}. The green curve corresponds to the same data as the red curve without smearing. The straight green line indicates a $k^{-1.71 \pm 0.245}$ power-law fit of the green curve. 
The straight black line corresponds to a $k^{-1.47 \pm 0.193}$ power-law fit of the underlying portion of the dashed orange curve. 
The two vertical dashed lines indicate the location of the integral and Taylor scales. 
Spectra from observations and simulations have been scaled because the observations are not calibrated to an absolute value.}
\label{fig1}
\end{figure*}

Figure~\ref{fig1} displays power spectra of the observed intensity from the quiet-Sun IMaX, the SuFI data, and the MHD simulations.  
The green curve corresponds to the power spectrum of the unsmeared simulated intensity in the $300$ nm band. The corresponding smeared simulated curve is represented in red. 
The similarity of the smeared simulations and the SuFI data between the integral scale and $\approx 320$ km is particularly remarkable.

In the subgranular range, the power spectrum of the unsmeared simulated intensity is more strongly curved than the power spectrum of the simulated Doppler velocity (as seen in the previous section). We nevertheless performed a least-squares fit of this curve with a power-law-shaped function between the integral and Taylor scales. The resulting power law has an index of $-1.71 \pm 0.245$. The power spectrum of the unsmeared simulated intensity apparently does not follow a power law closely in the region under consideration.
A power-law fit of the observed SuFI power spectrum exhibits an index of $-1.47 \pm 0.193$. The portion of the SuFI power spectrum under consideration appears to be reasonably well fit by a power law. To a somewhat lesser extent, this is also the case for the smeared simulations. 
The reason seems unclear, but it might be (in the case of the smeared simulations) due to a reminiscent almost strait part of the unsmeared simulations at the same wavelength range.
We cannot explain the case of the SuFI data clearly. 
The intensity is related to the velocity in an intricate way, therefore the appearance of a given power-law index in velocity does not simply predict one expected power-law index for intensity power spectra. 
A theoretical analysis involving both velocity and intensity might therefore be of great interest in order to better understand the nature of photospheric flows \citep[see, e.g.,][]{rieutord2010}.

In the literature, power-law indices of $-1$ \citep{Goode_etal_2010} and $-3$ and $-17/3$ \citep{rieutord2010} have been found from different datasets. This clearly keeps the debate still open.
Nevertheless, the agreement between the high-resolution SuFI observations and the smeared MHD simulations indicates that the simulations are likely to have captured some essential physical processes ongoing in the photosphere.

\section{Power spectra at different heights}

The simulations allow following the velocity power spectra from the optically thick solar interior into the optically thin atmosphere. Because these layers
are highly stratified (e.g., the density drops by more than $\approx 200$ times over a 1.4 Mm height range), the relative importance
of kinetic, internal, magnetic, and radiative energy transport also varies strongly  (e.g., the opacity drops by several orders of magnitude
over the same 1.4 Mm). We therefore expect that the properties of the turbulence are also strongly dependent on height.

\begin{figure}
\includegraphics[width=0.48\textwidth]{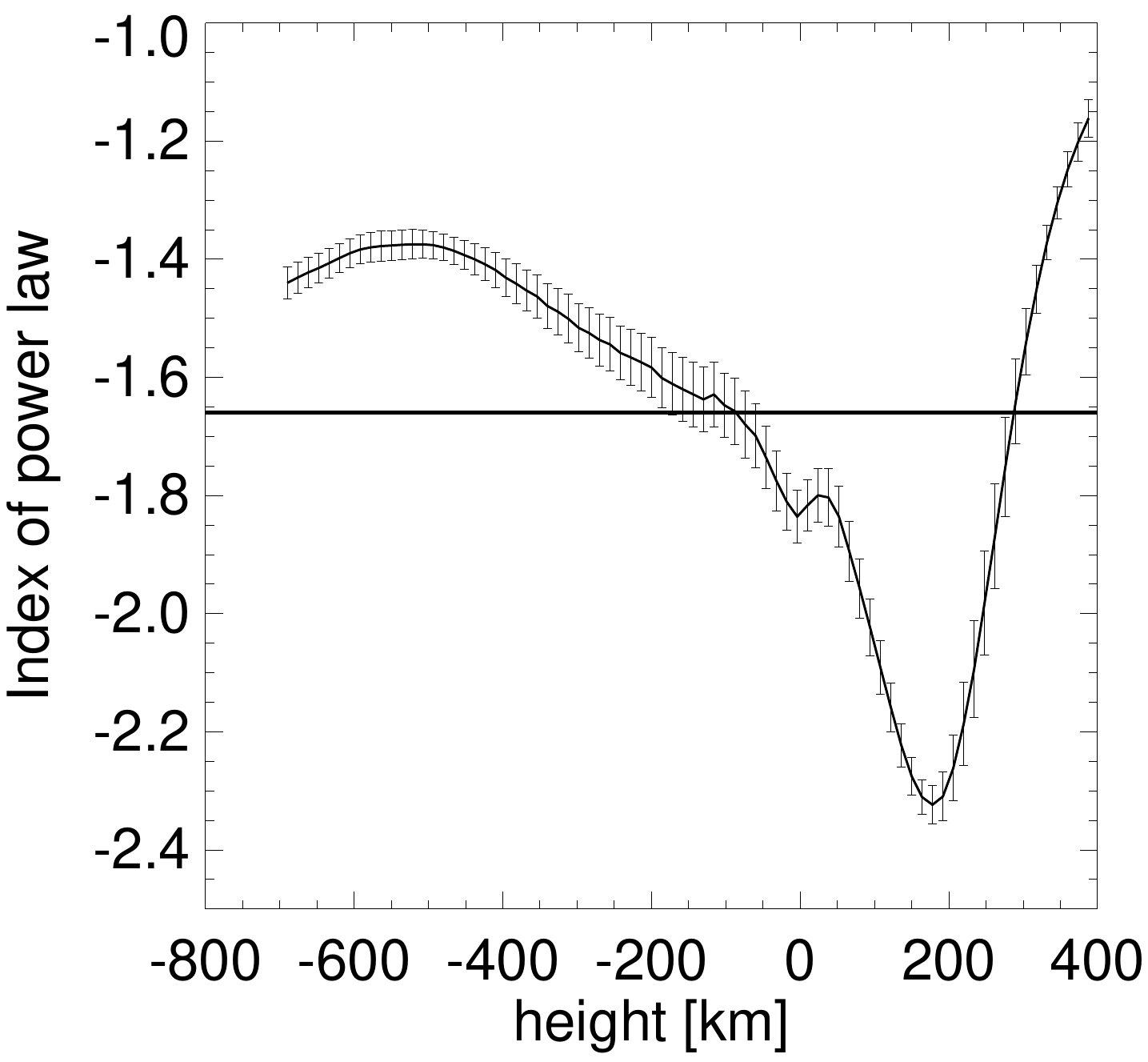}
\caption{Height dependence of the power-law indices of the vertical velocity power spectra
throughout almost the entire height range of the simulation box (the top and bottom parts, which may be affected by the boundary conditions, are not covered). A height of $0$~km indicates the location of the solar surface ($\tau_{5000}=1$). The horizontal line marks the Kolmogorov value of $-5/3$. The length of the vertical error bars represents the $\pm$ uncertainties of the least-squares fit to determine the power-law indices at each height in the numerical grid.}
\label{fig3}
\end{figure}

Figure~\ref{fig3} shows the power-law index in the subgranular range of vertical velocity power spectra obtained from simulations versus height.
The power spectra were computed for 2D maps of the vertical velocity taken at each position in the vertical direction in the numerical grid. We performed a least-squares fitting procedure with a power law between the integral and Taylor scales, as described in the previous sections.
In the lower part of the simulation box
(from $\approx -700$ km to $\approx -100$ km), the power-law index takes values of approximately $(-1.3$ to $-1.6)$ and is typically higher than the $-5/3$ Kolmogorov value.
As pointed out in the literature \citep[e.g.,][]{rieutord2010}, there are many differences between the homogeneous, isotropic turbulence (characterizing the classical Kolmogorov turbulence) and the more complex, inhomogeneous and compressible solar turbulence. Consequently, there is no particular reason to expect a $-5/3$ power-law index in these conditions. 
Higher up in the atmosphere (from $\approx 0$ km to $\approx 180$ km), the power spectra gradually steepen and reach a value of $\approx -2.32$ at a height of nearly $180$ km above the solar surface. Above this height, the power-law index steeply tends upward to less negative values and reaches subphotospheric values at a height of about $300$ km. Near the top of the computational domain (above $\approx 300$ km), the power-law index continues to increase as a result of the magnetic pressure and tension forces, which play an important role in the momentum equation (see Sect.~\ref{section_61}). 

Our main interest in this paper is the region between $\approx 0$ km and $\approx 300$ km. We therefore do not discuss the other regions in detail. The obtained range of
values of the power-law index throughout the simulation box is a challenge for classical convection theories (e.g., Bolgiano-Obukhov turbulence), which do not predict such a range of values for the slope
in a given convection setup \citep[see also the discussion in][]{rincon2006}. In order to begin to
understand the complicated physics in this region, we analyze the energy and momentum equations.

\subsection{Relative importance of the different terms in the energy equation}\label{section_51}

The energy equation employed in the simulations reads
\begin{eqnarray}\label{eq:velocityspectrum_and_kineticenergy}
  \frac{\partial \, e }{\partial t}\, + \nabla \cdot \left[ \textbf{\textit{v}} \left( e + p + \frac{\textit{B}^2}{8 \pi} \right) -\frac{1}{4 \pi} \textbf{\textit{B}} \left(\textbf{\textit{v}} \cdot \textbf{\textit{B}}\right)  \right] \nonumber\\
        = \frac{1}{4 \pi} \nabla \cdot  \left(\textbf{\textit{B}} \times \eta \nabla \times \textbf{\textit{B}} \right) + \nabla \cdot (\textbf{\textit{v}} \cdot \underline{\underline{\sigma}}) + \nabla \cdot (K \nabla T) \nonumber\\
        + \rho (\textbf{\textit{g}} \cdot \textbf{\textit{v}}) + Q_{rad} \;,
\end{eqnarray}
where $e = \rho\,\epsilon +0.5 \, \rho \textit{v}^2 + \textit{B}^2/8 \pi$ is the sum of the internal, kinetic, and magnetic energy densities per unit volume. $\textbf{\textit{v}}$ is the fluid velocity, $\textbf{\textit{B}}$ is the magnetic flux density, and $p$ is the gas pressure, $\rho$ is the mass density. $\underline{\underline{\sigma}}$ is the viscous stress tensor,
$T$ is the temperature, $K$ is the thermal conductivity, $\eta$ is the magnetic diffusivity, $\textbf{\textit{g}}$ is the gravitational acceleration, and $Q_{rad}$ is the frequency-integrated
radiative heating.   

\begin{figure}
\includegraphics[width=0.48\textwidth]{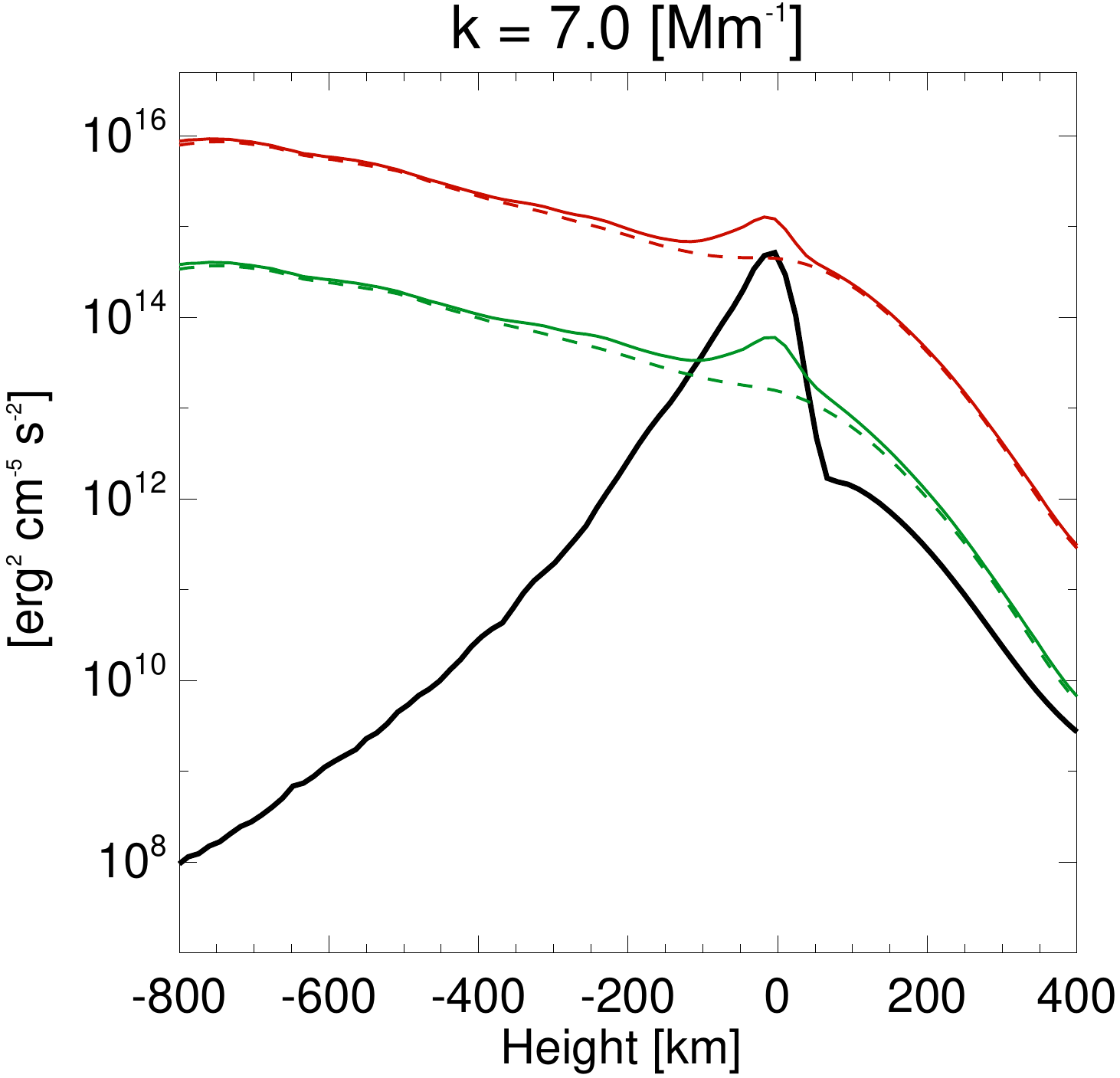}
\includegraphics[width=0.48\textwidth]{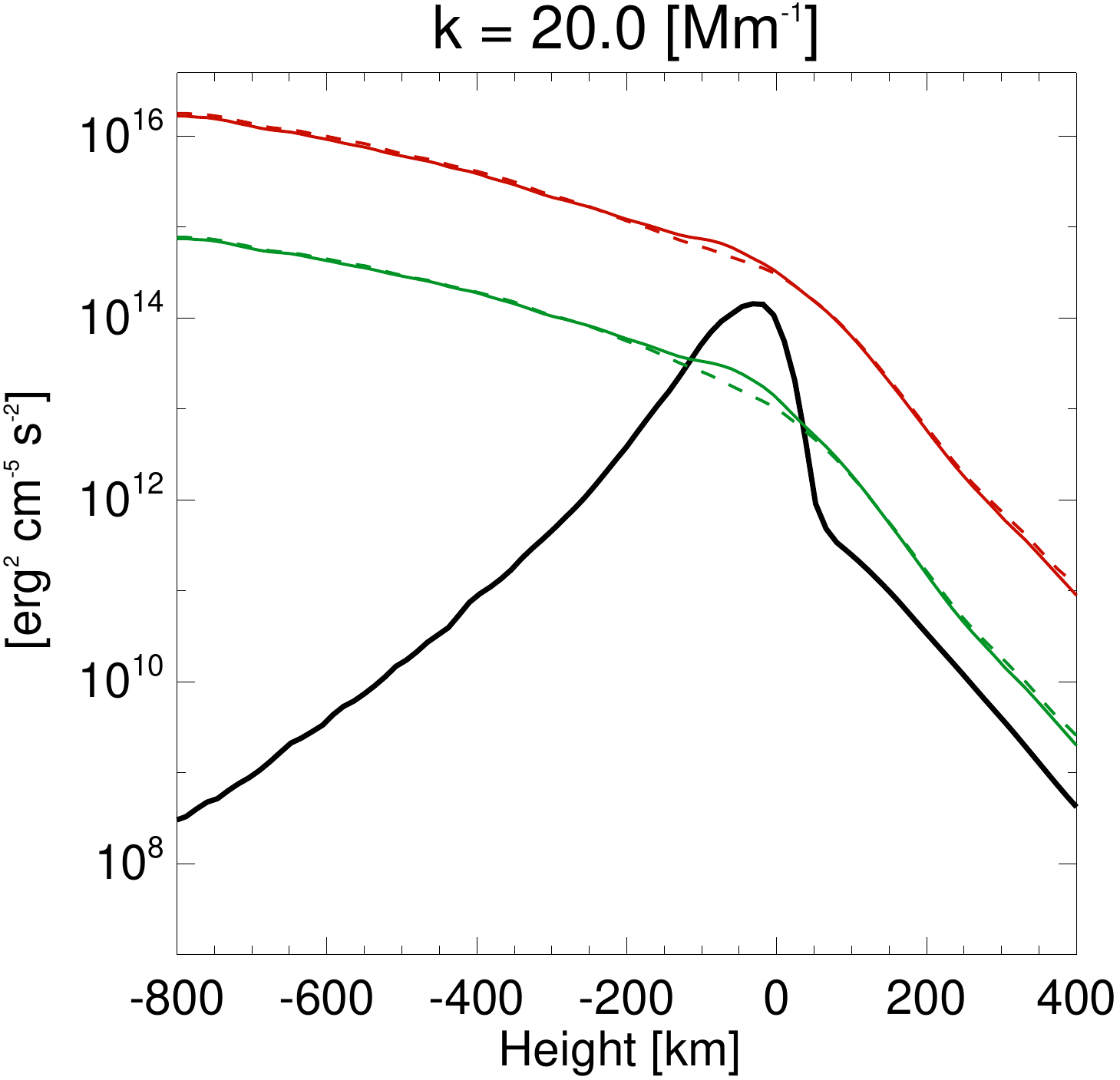}
\caption{Height dependence of the different terms in the energy transport at a wavenumber of $k=7$~Mm$^{-1}$, corresponding to a wavelength of
approximately $900$~km, in the upper panel and $k=20$~Mm$^{-1}$, corresponding to a wavelength of
approximately 312~km, in the lower panel. The solid black line shows the contribution from radiation, the solid (dashed) red line
the vertical (horizontal) transport of the internal energy, and the solid (dashed) green curve the work done  by the vertical (horizontal) pressure gradients.
}\label{fig_ps_energy}
\end{figure}

Figure~\ref{fig_ps_energy} shows the contribution of the most significant terms of this equation versus height
for two values of the wavenumbers in the simulations. These two values of k are chosen near the two ends of the range indicated by vertical dashed lines in Fig.~\ref{fig0}. The plots correspond to the power contained in the corresponding term of the energy equation, that is, as usually defined, it is proportional to the square of its Fourier transform with an appropriate normalization \citep[see, e.g.,][]{rieutord2010}.
For both wavenumbers, radiation becomes important at about $z=-100$~km, and becomes less important again near $z=50$~km. Below $z=-100$~km, the high opacity hinders
radiation from transporting energy. At $z=0$, where $\tau_{5000}=1$ on average, radiation plays a much stronger role. The dynamics in this
layer is highly nonadiabatic; the plasma exchanges energy with the radiation field. At this height, the stratification is
strongly subadiabatic, and the flow is dominated by overshooting convection.

\subsection{Relative importance of the different terms in the momentum equation}\label{section_61}

The momentum equation is 

\begin{eqnarray}\label{eq:momentum_equation}
  \rho \frac{\partial \, v }{\partial t}\, + \rho \left(v \cdot \nabla \right) v  = - \nabla p  + \frac{1}{4 \pi}(\nabla \times \textbf{\textit{B}}) \times \textbf{\textit{B}} + \rho \textbf{\textit{g}}    + \nabla \cdot \underline{\underline{\sigma}} \;.
\end{eqnarray}

\begin{figure*}
\centering
\includegraphics[width=0.48\textwidth]{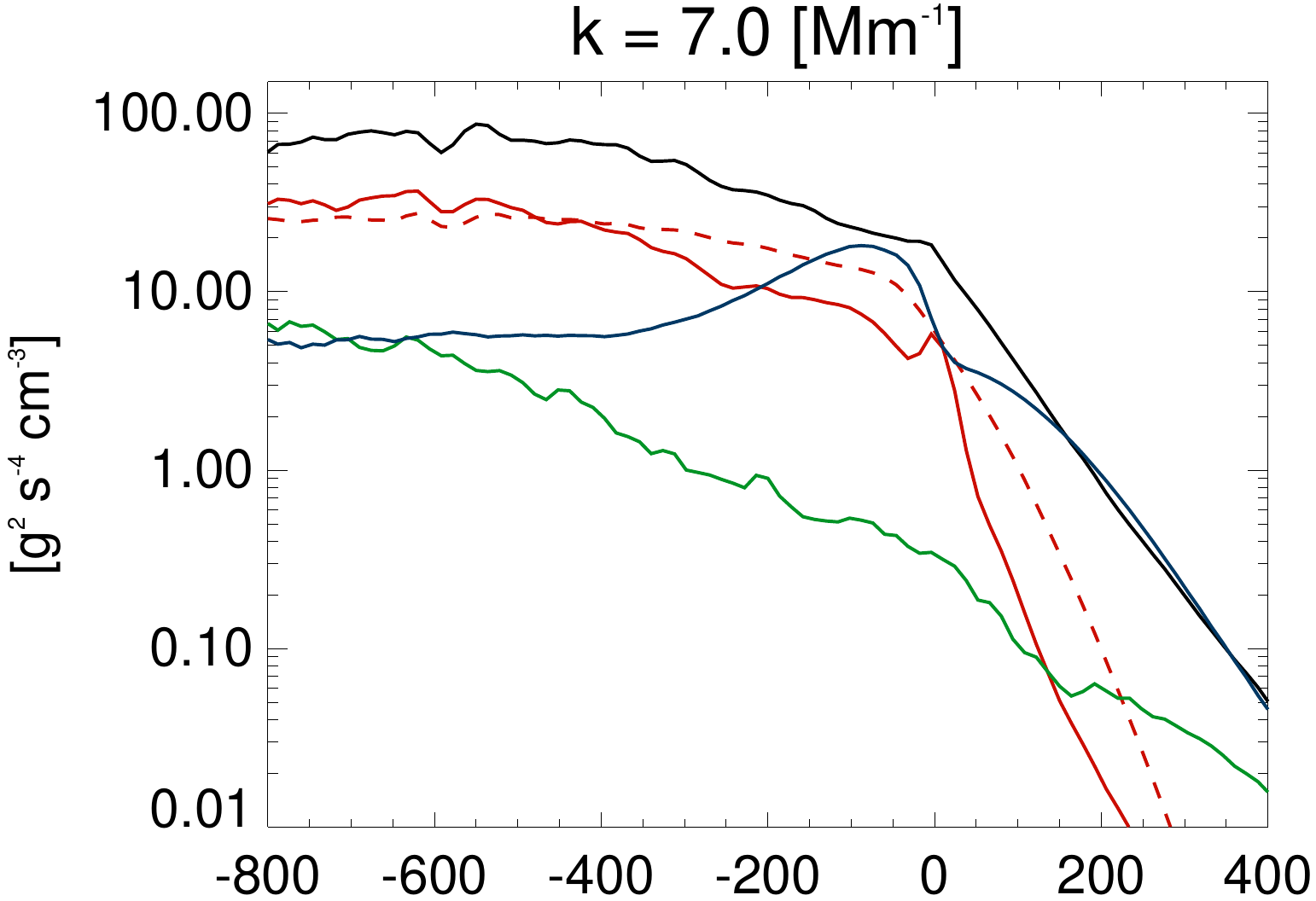}
\includegraphics[width=0.48\textwidth]{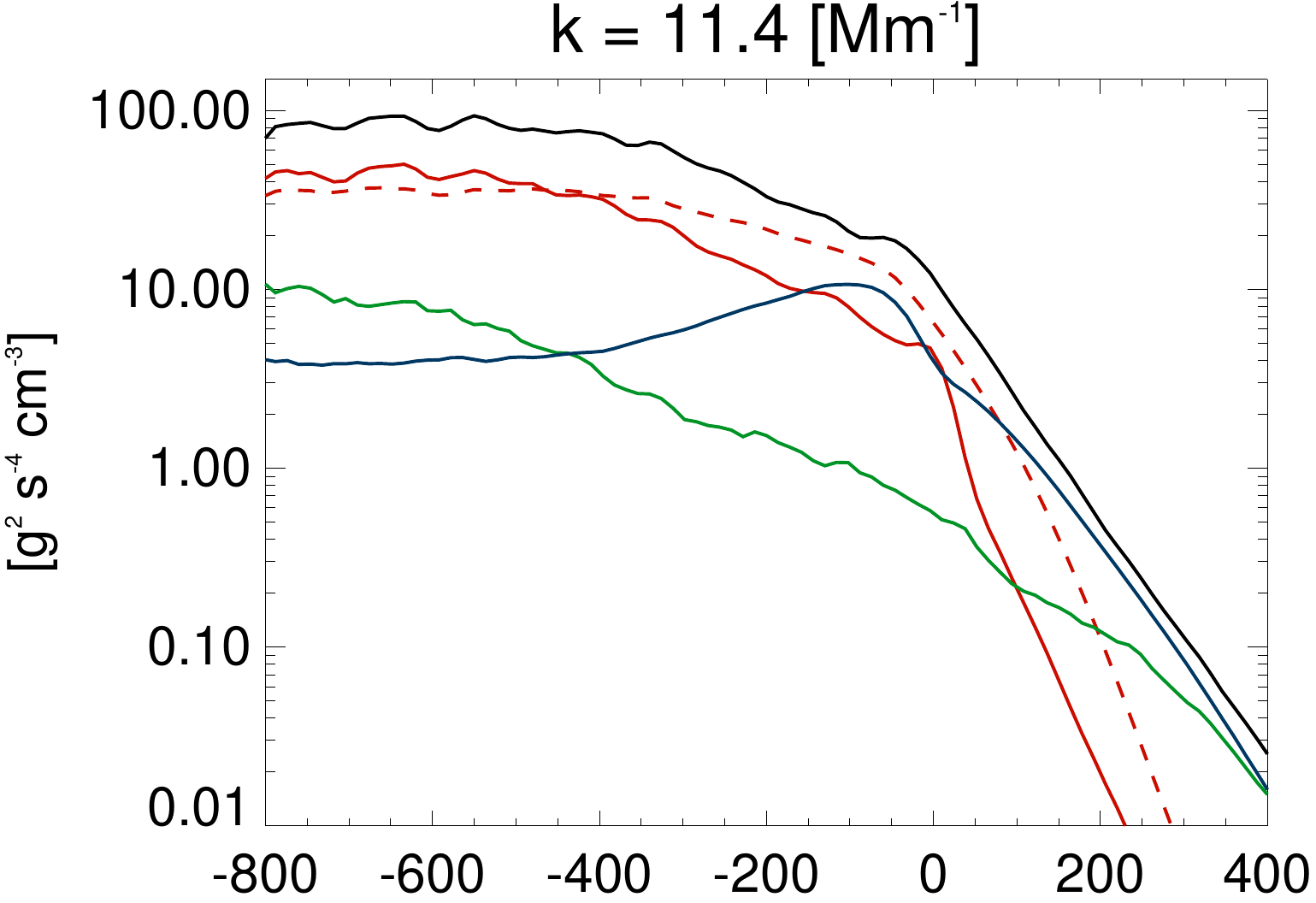}

\medskip
\centering
\vspace{1pt}
\includegraphics[width=0.48\textwidth]{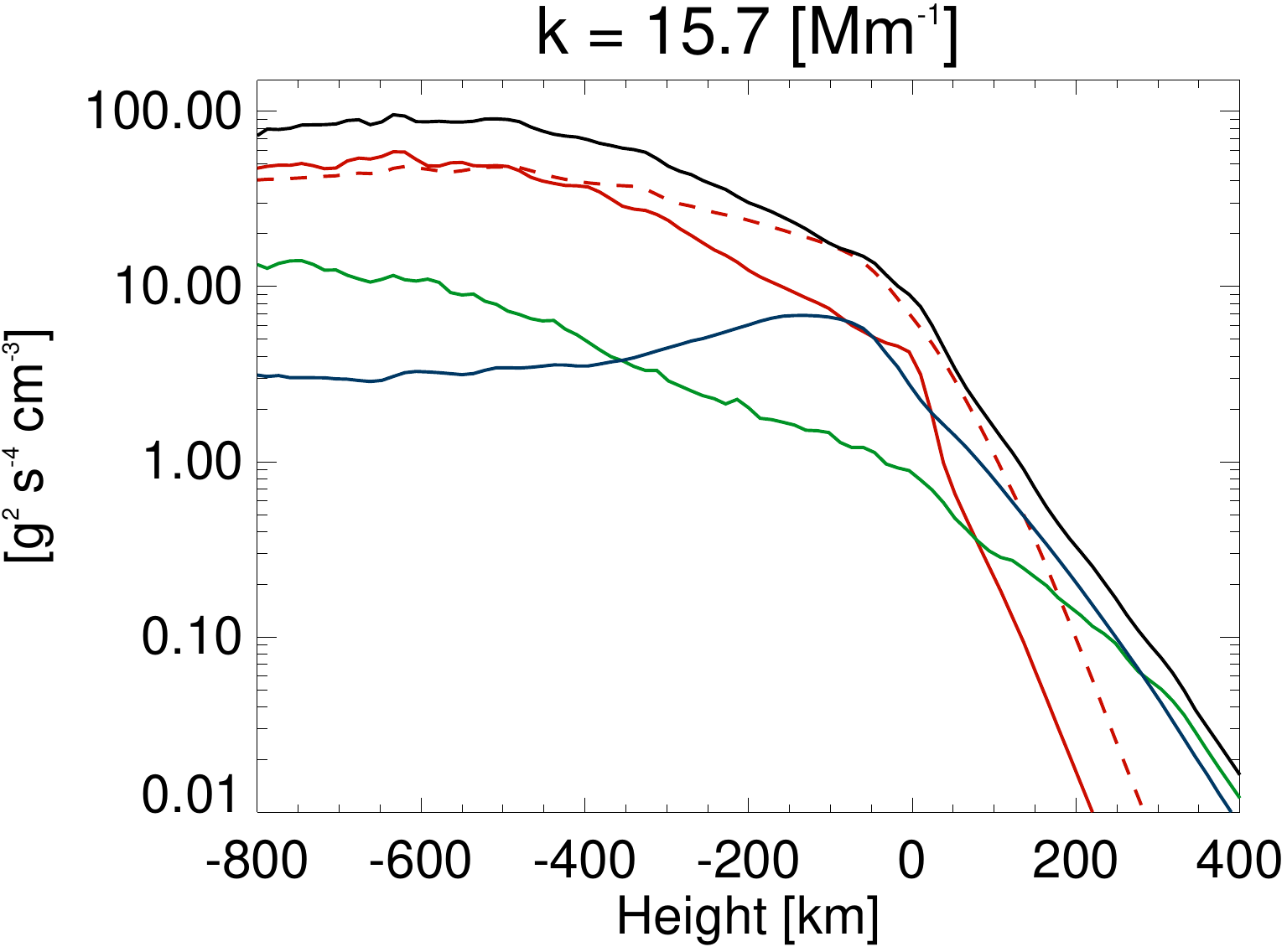}
\includegraphics[width=0.48\textwidth]{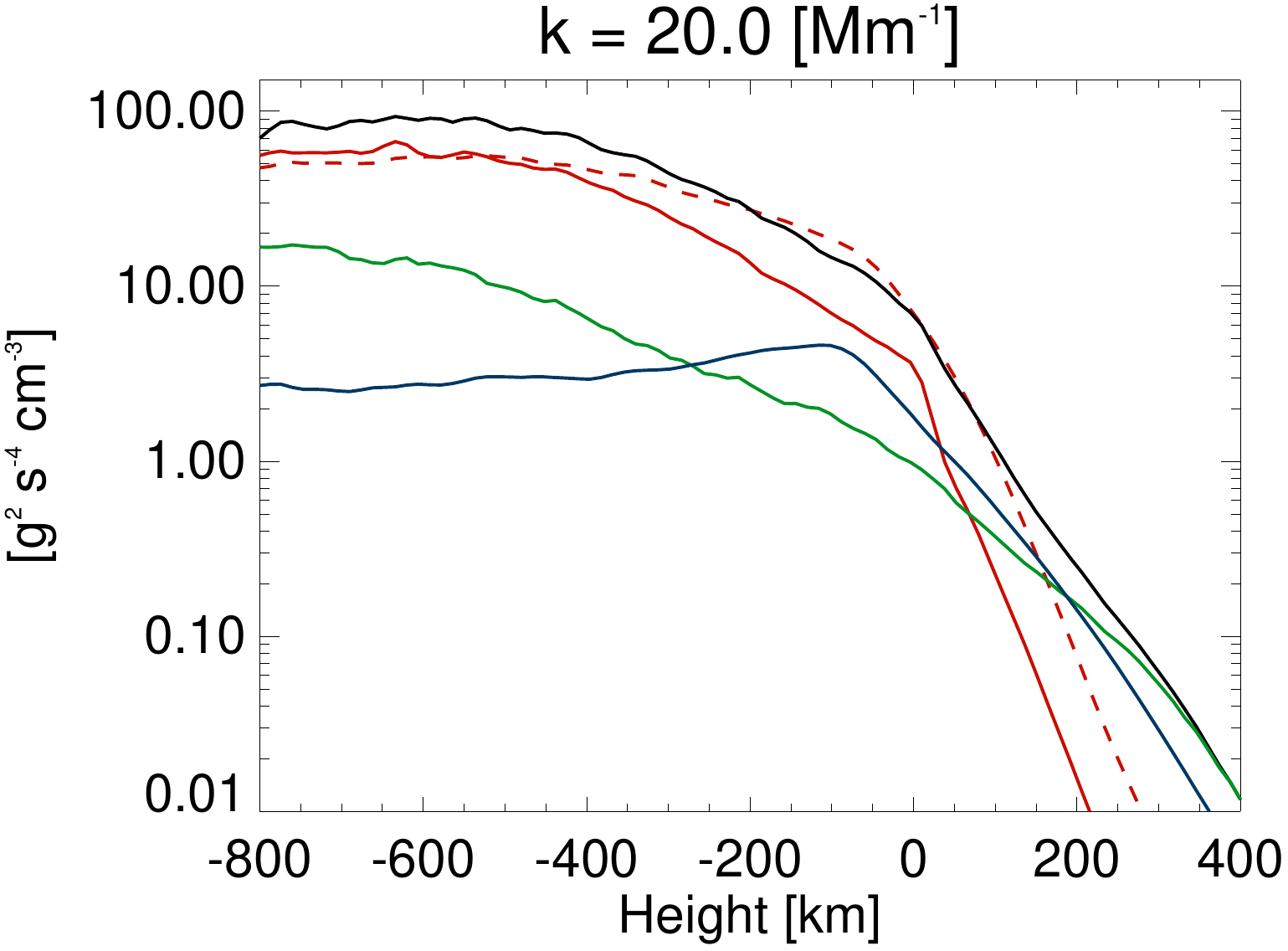}

\caption{Similar to Fig.~\ref{fig_ps_energy}, but now for the vertical component of momentum.
  The panels show the vertical (horizontal) transport of vertical momentum, represented by the solid (dashed) red curves. The black curve indicates the contribution of the vertical pressure gradient, the blue curve the
  contribution of gravity, and the green curve the contribution of the Lorentz force.
The panels from top left to bottom right are sorted from smaller to larger wavenumber. Top left: Results at wavenumber $k=7$~Mm$^{-1}$ (wavelength $900$~km). Top right: Wavenumber $k=11.4$~Mm$^{-1}$ (wavelength $550$~km). Bottom left: Wavenumber $k=15.7$~Mm$^{-1}$ (wavelength $400$~km). Bottom right: Wavenumber $k=20$~Mm$^{-1}$ (wavelength $312$~km).  
}\label{fig_ps_momentum}
\end{figure*}

It describes the evolution of the vector momentum. We focus on the vertical component of the momentum below. 
We considered the relative importance of the terms determining the evolution of the vertical component of the momentum, plotted in
Fig.~\ref{fig_ps_momentum}. As in Fig.~\ref{fig_ps_energy}, the plotted quantities are the power contained in the most relevant terms. 
Owing to the apparently complex physics involved in Fig.~\ref{fig_ps_momentum}, we focus first on the 
region that begins just above the solar surface and extends to $\approx 180$ km above it.
In order to clarify the origin of the steepening power-law indices in Fig.~\ref{fig3} (from the solar surface to almost $180$ km above it), 
three ingredients are relevant in Fig.~\ref{fig_ps_momentum}. First, the relative importance between the 
pressure gradient term and the horizontal transport of the vertical momentum term is scale dependent. 
The smaller the scale, the higher the relative contribution of the horizontal transport term (as shown in 
the four panels of Fig.~\ref{fig_ps_momentum}).
Second, the relative difference between the horizontal and the vertical transport terms is small 
near the bottom of the simulation box and become larger at heights corresponding to a power-law index 
steepening (especially above the solar surface).
Third, the work done by the gravity term is also scale dependent, and is larger for the larger scales.
%

Let us consider a plasma parcel that reaches the solar surface and travels into the subadiabatic atmosphere. If the horizontal size of that parcel is on the order of a granule (comparable to the integral scale), the pressure gradient is mainly compensated for by the work due to the gravitational field. This produces a break in the plasma motion. The atmosphere above the surface is almost hydrostatic at this scale, as seen by the dominance of the gravitational and pressure gradient contributions in the upper left panel of fig.~\ref{fig_ps_momentum} at a wavenumber of $900$ km. A plasma parcel of granular size can hardly be transported horizontally because it is surrounded by other granules. This is seen by the less important horizontal transport term (top left panel). If in contrast we consider a plasma parcel of a much smaller horizontal size (e.g., 312 km;  lower right panel of Fig.~\ref{fig_ps_momentum}), the vertical momentum is very efficiently transported horizontally, whereas gravity does not play a dominant role. 
For intermediate-size plasma parcels, the situation is between the two cases described above.  
As convection overshoots, the plasma parcels are therefore evacuated sideways with an efficiency that is inversely proportional to their horizontal size. 
Eventually, in the lower right panel, the pressure gradient is almost entirely balanced by the horizontal transport of the vertical momentum. This leads to a drastic lowering of the transport in the vertical direction, leaving less \textit{\textup{vertical momentum power}} to be transported to the atmospheric slice above. This process is progressively milder for increasingly larger scales (as shown in Fig.~\ref{fig_ps_momentum}).  
This means that starting at the solar surface and going upwards, a given scale distribution of vertical velocity that results in a given power-law index is therefore affected by the scale-dependent horizontal depletion of the vertical velocity. Each successive atmospheric slice (like in a numerical grid) will produce an increasaingly larger difference between the power near the integral scale and the power near the Taylor scale. This leads to progressively steeper power-law indices as the plasma travels from the solar surface to $\approx 180$ km as in Fig.~\ref{fig3}.  

The scale dependence of the relative importance between the pressure gradient and the horizontal and vertical transport terms also occurs below the solar surface (but in a less prominent way than above the solar surface) between $\approx -500$ km and the solar surface. It indeed produces a milder steepening of the power-law indices as well (Fig.~\ref{fig3}). 
Above $\approx 180$ km, most of the convectively transported material from below has already overturned, due to the relatively small pressure scale height. The work done by gravity takes over the horizontal transport term at all relevant scales. This drives the power-law indices to progressively return to flatter vales because the steepening mechanism described above is progressively less at work here. The plasma becomes more hydrostatic with height.       
At smaller wavelengths, as shown in Fig.~\ref{fig_ps_momentum} (bottom right panel) and near the top of the simulation box (above $~300$ km), the pressure gradients are balanced by the magnetic pressure and tension forces. On the one hand, the plasma beta becomes smaller near the top of the box. On the other hand, the upper boundary condition forces the magnetic field to become vertical at the top of the simulation domain, leading to tension forces due to the bending of the field lines. At greater heights, shocks become important \citep[][]{moll2012} 

\section{Conclusions}

We have presented power spectra observed at one of the highest spatial resolutions so far.
In the subgranular range, we found excellent agreement between the power spectra obtained from \sunrise\ observations (from the two instruments SuFI and IMaX)
and from advanced MHD simulations.
In the case of the Doppler velocity, the obtained power laws are significantly flatter than most of those obtained in the past.
The instrumental degradation inherent to the observational conditions of \sunrise\ reduces the extent of the power-law-shaped part of the power spectra compared to the case of the original simulations. This suggests that at lower resolution, the power-law-shaped part eventually has an even smaller extent. This might result in larger uncertainties in the determination of the corresponding power-law exponent. With a view to better determining the photospheric power spectra and the physical processes related to them, it therefore appears necessary to observationally reach higher wavenumbers.

The power-law indices of the vertical velocity power spectra depend sensitively on the height in the atmosphere,
as deduced from MHD simulations, because the motions we observe in the lower photosphere are dominated by overshooting convection,
which penetrates to heights that depend on their horizontal wavelengths. 
The nonadiabatic damping of the convective motions produces a scale-dependent transport of the vertical momentum. This in turn produces a gradual steepening of the power-law indices between the solar surface and $\approx 180$ km above. At still greater heights, most of the plasma coming from below has already overturned. Gravity work starts taking over transport terms in the momentum equation at all relevant scales, leading to progressively flatter power-law indices.

Our findings show that
simulations and high-resolution observations agree and together
reveal some properties of the convectively driven flows in the solar photosphere and subphotospheric layers. 
Nevertheless, many aspects remain to be understood, such as the implications of a given intensity power spectrum on the plasma properties and its link to the velocity spectra \citep[e.g.,][]{rieutord2010}. On the observational side, advances (as in the case of the Daniel K. Inouye Solar telescope) will allow exploring the data up to higher wavenumbers with the view of providing further indications of the underlying physics in the solar context and possibly beyond.

\begin{acknowledgements}
The German contribution to \sunrise{} and its reflight was funded by the
Max Planck Foundation, the Strategic Innovations Fund of the President of the
Max Planck Society (MPG), DLR, and private donations by supporting members of
the Max Planck Society, which is gratefully acknowledged. The Spanish
contribution was funded by the Ministerio de Econom\'ia y Competitividad under
Projects ESP2013-47349-C6 and ESP2014-56169-C6, partially using European FEDER
funds. The HAO contribution was partly funded through NASA grant number
NNX13AE95G. This work was partly supported by the BK21 plus program through
the National Research Foundation (NRF) funded by the Ministry of Education of
Korea and by the European Research Council (ERC) under the European Unions Horizon 2020 research and innovation program (grand agreements No. 715947 and No. 695075). 
\end{acknowledgements}

\bibliographystyle{aa}
\bibliography{magnetic6}

\end{document}